\documentclass[manuscript]{acmart}

\usepackage[yyyymmdd]{datetime}

\usepackage{tabularx}
\usepackage{dcolumn} 
\newcolumntype{d}[1]{D{.}{.}{#1}}

\usepackage{subcaption}

\usepackage{todonotes}
\let\xtodo\todo
\renewcommand{\todo}[1]{\xtodo[inline,color=green!50]{#1}}

\usepackage{booktabs} 
\usepackage{siunitx} 
\usepackage{caption} 
\usepackage{graphicx} 
\usepackage{xcolor}
\usepackage{multirow}
\usepackage{colortbl}
\usepackage{xcolor}
\usepackage[normalem]{ulem} 
\definecolor{DeleteGrey}{RGB}{120,120,120}
\definecolor{AddColor}{RGB}{0,128,0}

\AtBeginDocument{%
  }

\setcopyright{none}
\renewcommand\footnotetextcopyrightpermission[1]{}
\acmISBN{978-1-4503-XXXX-X/18/06}

\begin{document}

\title{Physiological and Behavioral Modeling of Stress and Cognitive Load in Web-Based Question Answering}


\settopmatter{authorsperrow=3}

\author{Ailin Liu}
\orcid{0009-0002-7489-058X}
\affiliation{%
  \institution{LMU Munich}
  \city{Munich}
  \postcode{80337}
  \country{Germany}}

\author{Francesco Chiossi}
\orcid{0000-0003-2987-7634}
\affiliation{
  \institution{LMU Munich}
  \country{Germany}
}

\author{Felix Henninger}
\affiliation{
  \institution{LMU Munich}
  \country{Germany}
}

\author{Lisa Bondo Andersen}
\affiliation{
  \institution{LMU Munich}
  \country{Germany}
}

\author{Tobias Wistuba}
\affiliation{
  \institution{HU Berlin}
  \country{Germany}
}

\author{Sonja Greven}
\affiliation{
  \institution{HU Berlin}
  \country{Germany}
}

\author{Frauke Kreuter}
\affiliation{
  \institution{LMU Munich}
  \country{Germany}
}

\author{Fiona Draxler}
\orcid{0000-0002-3112-6015}
\affiliation{
  \institution{University of Mannheim}
  \country{Germany}
}

\renewcommand{\shortauthors}{Liu et al.}


\begin{abstract}
Time pressure and question difficulty can trigger stress and cognitive overload in web-based surveys, compromising data quality and user experience. Most stress detection methods are based on low-resolution self-reports, which are poorly suited for capturing fast, moment-to-moment changes during short online tasks. Addressing this gap, we conducted a 2×2 within-subjects study (N = 29), manipulating question difficulty and time pressure in a web-based multiple-choice task. Participants completed general knowledge and cognitive questions while we collected multimodal data: mouse dynamics, eye tracking, electrocardiogram, and electrodermal activity. Using condition-based and self-reported labels, we used statistical and machine learning models to model stress and question difficulty. Our results show distinct physiological and behavioral patterns within very short timeframes. This work demonstrates the feasibility of rapidly detecting cognitive-affective states in digital environments, paving the way for more adaptive, ethical, and user-aware survey interfaces.
\end{abstract}


\begin{teaserfigure}
  \includegraphics[width=1.0\textwidth]{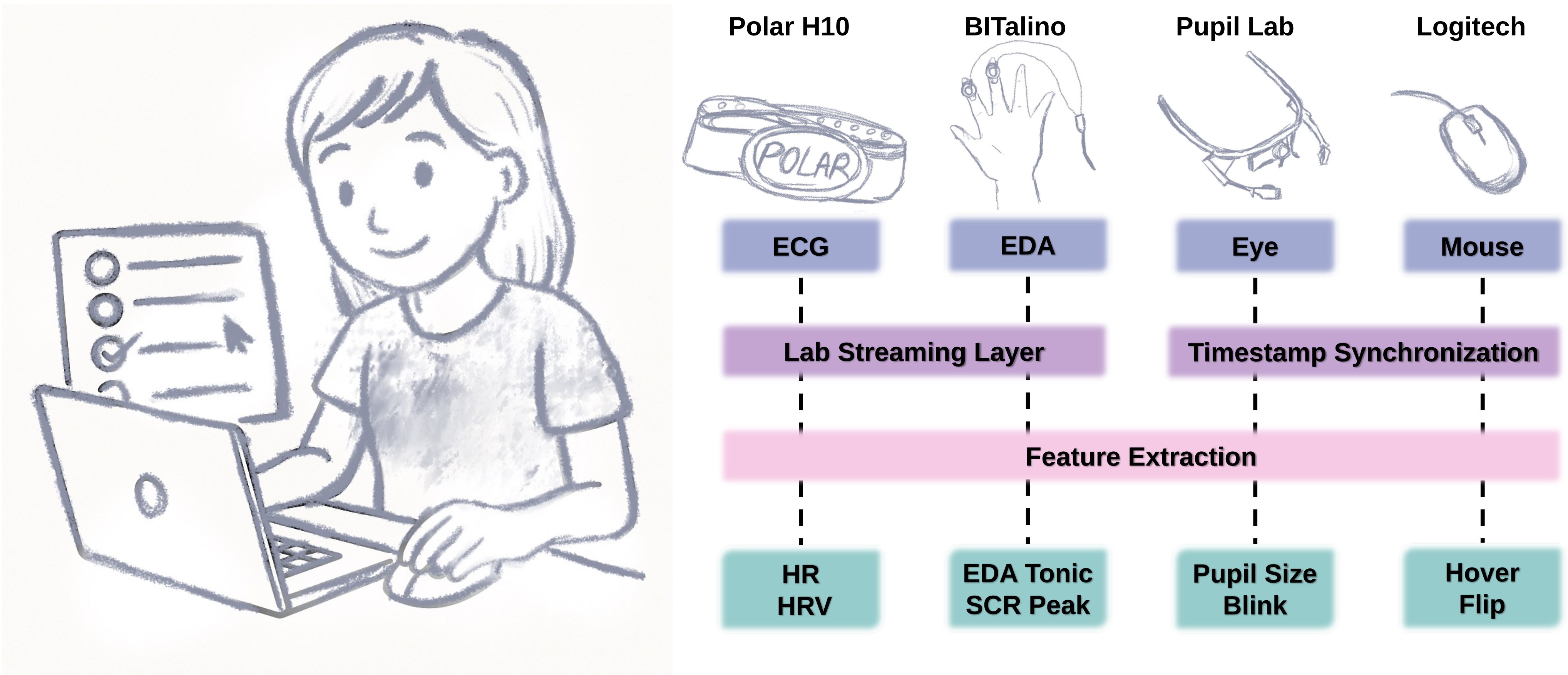} 
    \caption{Experimental setup integrating web-based task presentation with multimodal sensor data acquisition. The system simultaneously collected ECG, EDA, eye tracking, and mouse tracking data.}
    \label{fig:overview}
\end{teaserfigure}




\maketitle

\section{Introduction}

Online surveys have fundamentally transformed data collection across social science and Human–Computer Interaction (HCI) research. Web-based instruments offer substantial advantages over traditional methods: low-cost access to large samples, geographic flexibility, and interface features that can personalize respondent experience \cite{Callegaro2015}.Remote data collection has grown steadily across HCI research and practice, with analyses showing that a substantial share of HCI studies, around 20\% as early as 2016, were conducted remotely \cite{Caine2016}. As remote methods have become even more prevalent, general-population surveys are now frequently administered online. This expansion heightens longstanding concerns about measurement error, particularly when respondents misinterpret or struggle to answer survey questions \cite{Tourangeau2003, berry2022drivers, decieux2024sequential}.

Unlike interviewer-administered surveys, where trained professionals can detect confusion and offer clarification \cite{10.1086/269312}, self-administered web surveys provide no real-time human support. When respondents encounter cognitively demanding items or feel pressured by time constraints, they may engage in satisficing behaviors due to high workload, selecting suboptimal responses to reduce effort, which compromises data quality \cite{Blazek2023, Krosnick1991}. Survey methodologists have therefore emphasized the need to detect problematic items, whether through pretesting before deployment, real-time intervention during administration, or post-hoc identification of unreliable responses \cite{Matjai2024, Schneider2024, https://doi.org/10.18148/srm/2024.v18i3.8304}.

Traditional approaches to improving web survey quality rely on static design decisions made before any respondent interaction occurs. Pretesting techniques identify and revise difficult questions \cite{demaio1998improving} and standardized formats aim to reduce universal cognitive burden \cite{Lenzner2009, Brosnan2021, Kunz2024}. More recently, paradata and behavioral traces such as response times and mouse movements, have enabled post-hoc detection of problematic items \cite{Lenzner2009, Matjai2024, Schneider2024, https://doi.org/10.18148/srm/2024.v18i3.8304}. Cursor-tracking techniques, for instance, reveal hesitation or uncertainty through movement patterns, offering retrospective indicators of question quality \cite{leipold2024detecting, fernandez2023predicting}.

Despite their value, these techniques share a fundamental limitation: they cannot response to or intervene individual respondent struggles as they occur, meeting personal needs. A question labeled “difficult” in pretesting may be trivial for domain experts yet overwhelming for novices; similarly, time pressure may improve performance for some respondents but induce anxiety in others, depending on the individual’s stress mindset \cite{Zhou2024}. Static personalization, based on pre-determined categories such as education level, cannot account for moment-to-moment fluctuations in cognitive load, emotional state, or task-specific knowledge that shape each individual’s survey experience \cite{simonsmeier2022domain, judd2024interindividual}.

Recent advances in ubiquitous sensing and real-time inference have enabled adaptive systems in domains such as driving or workplace tasks \cite{schaule2018employing, park2024hide, luvstrek2023designing, fridman2018cognitive, balters2020calm}. These systems infer cognitive–affective states, including distraction, stress, and mental overload, from multimodal signals such as electrodermal activity (EDA), heart rate (HR), heart rate variability (HRV), eye and mouse dynamics. Upon detecting difficulty, adaptive interfaces adjust task presentation by offering hints, reducing complexity, or extending time limits. For web surveys, similar capabilities could transform error prevention: rather than retrospectively identifying problematic responses, systems could offer clarification when mouse trajectories indicate confusion, remind respondents to relieve stress when physiological signals indicate time–induced stress, or provide rest breaks when cumulative cognitive load becomes apparent.

Realizing this vision requires overcoming a unique set of challenges. First, survey questions involve brief interactions, often 5–15 seconds, requiring faster detection than systems designed for sustained tasks \cite{schaule2018employing, park2024hide, luvstrek2023designing, fridman2018cognitive, balters2020calm}. Second, individual variation in cognitive appraisal \cite{Lazarus1991} and knowledge means that objective task properties may poorly predict subjective difficulty \cite{Kunz2024}, a discrepancy that has rarely been examined in adaptive system research.

These challenges expose a deeper theoretical gap. Surveys involve three distinct forms of task-related factors:
(1) Design-based task characteristics, such as manipulated difficulty or time pressure, reflecting the external conditions the researcher intends to create;
(2) Self-reported cognitive load and stress, reflecting participants’ internal experience; and
(3) Multimodal behavioral and physiological signals, reflecting the observable traces of that experience.

Rather than treating either design or self-report as the single valid ground truth, we adopt an interactionist perspective, in which user state emerges from the dynamic coupling of task design and personal experience. Accordingly, our goal is not to compare signals against a single ground truth, but to characterize the alignment
among task design, subjective experience, and multimodal signals. Specifically, we are interested in (1) how manipulated task properties shape behavioral and physiological responses (a dimension from external design to observed signals); (2) how self-reported cognitive load and stress manifest in these signals (a dimension from internal experience to observed signals); and (3) how psychophysiological signals inform the design of adaptive survey systems (a dimension from observed signals to adaptive systems). 

We address this gap through a controlled laboratory 
study (N=29) that systematically manipulates question difficulty and time pressure while 
collecting multimodal data: mouse dynamics, eye tracking, EDA, and electrocardiogram (ECG). Critically, we compare models on experimentally manipulated conditions versus participants' immediate self-reported stress and difficulty to investigate the relationships among given task demands, self-experience, and psychophysiological signals on tasks. This design reveals both the feasibility of rapid  detection and the extent to which objective conditions diverge from subjective experience, informing whether adaptive systems should respond to task characteristics, user behaviors, or some combination. Our findings demonstrate three key contributions:
\begin{enumerate}
  \item \textbf{Methodological Implication:} We show that effective adaptive support in brief interactions like survey questions should consider jointly objective task design and subjective experience, as cognitive appraisal introduces variance that objective manipulations alone cannot capture.
  \item \textbf{Empirical Foundation:} We identify which physiological and behavioral signals are sensitive to specific cognitive–affective demands within sub-20-second windows, which extends adaptive sensing from continuous tasks to discrete survey items.
  \item \textbf{Design and Practical Implication:} Based on the observed mappings between design, experience, and signals, we propose a tiered micro-intervention framework outlining when and how signals can trigger real-time adjustments (e.g., clarification, timing changes, brief breaks). This framework is broadly applicable to remote HCI and social-science studies that rely on screen-based interactions and short, task-oriented activities.
\end{enumerate}

\section{Background and Related Work}
Cognitive load and time stress are central factors shaping how respondents process questions and provide answers in surveys. Cognitive load reflects the mental effort required to interpret and respond to survey items, while time stress arises when respondents perceive pressure to answer quickly, often affecting response quality and engagement. This section contextualizes mental stress and cognitive load in survey interactions, reviewing prior work on: (1) cognitive load and stress in HCI and affective computing, (2) measuring and modeling users' cognitive-affective states, (3) data challenges and innovations in web-based surveys, and (4) the relationship between task demand and users' lived experience.

\subsection{Cognitive Load and Stress in HCI and Affective Computing}
Cognitive load and stress are central constructs for understanding how users engage with interactive systems. Within HCI, these factors have been extensively examined across diverse domains, including learning interfaces \cite{Sharmin2025, Cherif2025, Kazemitabaar2025, Gebreegziabher2025}, complex decision-making environments \cite{Cao2023, Manzano2021, Zhou2017}, immersive technologies \cite{Fang2025, Pata2025, Chiossi2025}, and productivity tools \cite{Gadhvi2025, DeLaTorre2025, Brun2025, Knierim2025}. Across these contexts, mental effort and affective strain are key determinants of user performance, engagement, and satisfaction \cite{Niwanputri2025, Wang2025, Schmutz2009, Sharmin2025}.

Affective computing extends this perspective by conceptualizing cognitive and emotional states as integral components of interaction that can inform adaptive system behavior \cite{picard2000affective, Maybury1998}. In this view, detecting affective states in situ enables intelligent adaptation, where interfaces dynamically adjust their interaction style to align with users’ goals, attitudes, plans, and capabilities. Foundational models such as Picard’s affective loop \cite{picard2000affective} and McDuff’s user state adaptation framework \cite{McDuff2012} illustrate how sensing user affect can guide system adjustments in feedback, pacing, or task difficulty.

Adaptive and intelligent HCI systems aim to adjust interaction demands to users’ cognitive and affective states. Designing such systems requires considering not only the user interface but also the task demands and contextual constraints that shape how much mental effort a user must expend \cite{saffer2010designing}. In this context, workload is a key construct: it reflects both the amount of information a user must process and the time available to process it—factors that jointly influence cognitive load and perceived stress \cite{szalma2007task, Hancock1989, Bong2016, Cao2023, Aigrain2018}. This perspective positions cognitive- and affect-sensing as essential to enabling interfaces that support efficient, comfortable, and reliable interaction \cite{Karray2008, chiossi2022adapting}. In a survey context, question difficulty and the time respondents need to complete the questionnaire are also common survey-design measures of objective burden \cite{Galesic2009, Rolstad2011, Kunz2024}. Consequently, detecting signs of cognitive overload and stress within this adaptive framework is essential for enabling intelligent interactive systems to anticipate user strain and dynamically adjust interaction demands in real time.

\subsection{Measuring and Modeling Users' Cognitive-Affective States}

Measurement of users’ cognitive-affective states has traditionally relied on two complementary approaches: subjective self-report and objective physiological assessment \cite{Kosch2023}. Subjective methods typically involve questionnaires or self-assessments that capture users’ perceived mental effort, stress, or affective state. While these tools provide direct access to subjective experience, they are limited by recall bias, lack of temporal granularity, and susceptibility to self-presentation effects. Objective methods, by contrast, employ non-invasive biometric sensors to collect continuous physiological data indicative of autonomic and cognitive activity \cite{Suzuki2023}. While these signals can provide more timely and fine-grained indicators than self-reports, they are also sensitive to individual differences and contextual factors. Among behavioral indicators, mouse tracking has gained attention for revealing real-time cognitive dynamics through cursor trajectories. Unlike response-time measures that capture only the duration of a decision, mouse trajectories reflect continuous motor–cognitive coupling \cite{cisek2010neural, freeman2010mousetracker, freeman2018doing}. Deviations in trajectory shape can indicate hesitation, conflict, or uncertainty \cite{cepeda2018mouse, daCSilva2023, Yamauchi2017}, providing a scalable, low-cost alternative to laboratory-based sensing methods such as eye tracking.

Parallel work on detecting cognitive load and stress increasingly leverages physiological signals including EDA \cite{nigam2021improved, zontone2019stress, Nourbakhsh2017}, eye tracking \cite{korda2021recognition, dao2024state, heimerl2022pupillometry, Heimerl2025}, and cardiovascular measures such as HR and HRV \cite{kim2018stress, immanuel2023heart, 10.1145/3340555.3353735, 10.1145/3453892.3461625, Haapalainen2010}. These signals provide temporally fine-grained and difficult-to-mask insights into the autonomic nervous system (ANS) \cite{barreto2007non, pinge2024detection}, enabling ecologically valid stress monitoring.

Existing empirical studies span a wide range of task durations and contexts. Longitudinal work such as \citet{Booth2022} examines stress and cognitive load across extended periods, while \citet{Heimerl2025} investigate high-stakes interactions such as job interviews lasting around 60 minutes. Other studies focus on specific cognitive experiences, such as narrative comprehension \cite{Zhang2025}, or everyday productivity tasks lasting 5–10 minutes \cite{daCSilva2023, https://doi.org/10.6084/m9.figshare.9970304}. Across these varied contexts, prior work consistently demonstrates the utility of psychophysiological measures in identifying and distinguishing users’ stress and cognitive overload. However, limited studies have examined extremely short-duration interactions, such as sub-20-second tasks, where cognitive and affective responses may unfold rapidly. Experimental stressors have also been introduced through time-constrained cognitive-test tasks \cite{Aigrain2018, Markova2019, Chen2021}, though these typically evaluate blocks of questions rather than very short interaction windows, i.e., a single question. As a result, cognitive–affective responses at fine temporal scales, for instance, within sub-20-second question intervals, remain underexplored, despite their importance for diagnosing and improving survey data quality.

Across these approaches, a central challenge for HCI is not only achieving accurate measurement but also developing interpretive models that explain how multimodal signals relate to users’ lived experience of interaction. Advancing this understanding requires moving beyond simple classification toward models that capture how task demands, appraisal processes \cite{Lazarus1991}, and cognitive–affective states evolve dynamically.

\subsection{Data Challenges and Innovations in Web-Based Survey}

Web-based surveys face persistent data-quality threats such as satisficing \cite{Blazek2023}, item nonresponse \cite{Haunberger2011}, dropout \cite{Hoerger2010}, straightlining \cite{https://doi.org/10.18148/srm/2014.v8i2.5453}, and automated bot responses \cite{Bonett2024, Caven2025}. Many of these behaviors arise when cognitive demands exceed respondents’ available cognitive resources \cite{Krosnick1991}. \citet{Krosnick1991} argues that respondents under cognitive strain may adopt satisficing strategies ranging from selecting the first acceptable option to random responding or item skipping. Even a small proportion of low-quality responses can bias correlations, factor structures, and group comparisons \cite{Huang2015, Schneider2017, Ward2023}. Yet much prior work attributes these issues primarily to low motivation, “careless responding” \cite{Meade2012}, while underemphasizing cognitive ability and cognitive-affective status as central contributors to degraded data quality.

To address these challenges, researchers increasingly use metadata and paradata, such as response times, to infer cognitive effort and detect problematic items or respondents \cite{Matjai2024, Schneider2024, https://doi.org/10.18148/srm/2024.v18i3.8304}. Although effective for post-hoc detection, these methods treat cognitive overload as a nuisance factor rather than a dynamic component of the interaction. Design innovations attempt to intervene earlier: for example, the two-phase Organize-then-Vote interface \cite{Cheng2025} encourages deeper engagement and reduce cognitive load in order to reduce satisficing, but remains one-size-fits-all and cannot adapt to individual differences in cognitive capacity. Gamification can increase completion rates \cite{Harms2014}, yet risks compromising construct validity. Systems-level approaches link survey items to external behavioral data \cite{Velykoivanenko2024} or provide platforms for interactive online studies \cite{Ebert2023}, but still do not address any existing data quality challenge.

Overall, prior innovations focus on detecting low-quality responses or redesigning survey interfaces, but rarely examine how physiological and behavioral signals could be leveraged proactively to prevent degradation. This gap highlights the need for an HCI perspective that treats cognitive load as an inherent part of the interaction. Our work advances this agenda by characterizing the alignment between designed task demands, experienced cognitive-affective states, and multimodal physiological and behavioral signals, offering foundations for more adaptive, user-aware survey systems.

\subsection{From Task Demand to Lived Experience}
Cognitive Appraisal Theory provides a foundational lens for linking users’ lived experiences with the physiological and behavioral signals that accompany interaction \cite{Folkman2013, Lazarus1991}. Rooted in cognitive psychology, this theory posits that emotion and stress emerge from individuals’ cognitive appraisals of internal and external events, shaped by prior knowledge and experience \cite{Lazarus1991, martin2013theories, szalma2007task, Saavedra2025}. In this framework, the meaning users assign to a situation determines their emotional response, which in turn manifests through measurable physiological and behavioral changes \cite{Bagozzi1999}. Cognitive Appraisal Theory thus encompasses three tightly coupled elements: the cognitive appraisal of external stimuli, the resulting affective reaction, and the behavioral or physiological responses that follow. The intensity and nature of these responses depend on how users evaluate the fit between their goals, expectations, and the system’s demands.

Because stress and cognitive strain arise from the dynamic interaction between individuals and their environments, understanding user experience requires examining how personal states and contextual factors jointly shape appraisal processes. Therefore, treating stress and cognitive load as isolated constructs neglects the systemic interdependencies between person and task \cite{szalma2007task, Zhang2024}. In other words, understanding the mapping between design parameters, experienced states, and multimodal signals can inform next-generation adaptive interfaces that infer user states in situ. A holistic alignment between design intent, user experience, and physiological and behaviroal response is therefore essential to advance adaptive and human-centered interaction design \cite{szalma2007task, Larradet2019}.

Recent work has begun to explore these appraisal processes within HCI. For example, Saavedra \cite{Saavedra2025} employed physiological signals to examine unconscious user responses to visual stimuli. Evidence also shows that stress is not a direct consequence of any single environmental factor, but rather an emergent appraisal of task demands relative to one’s perceived personal resources \cite{sandi2013stress, al2023essential}. However, little research has examined how everyday survey design choices, such as time pressure or task difficulty, shape users’ appraisals and the resulting multimodal stress and cognitive-load responses. This gap matters because appraisal theory provides a mechanism for understanding why users react differently to identical survey tasks: subjective evaluations may reflect affective predispositions rather than actual performance and be vulnerable by self-reflection biases or socially desirable responding \cite{Kosch2022, Natesan2016}. By contrast, objective indicators such as task completion time, error rates, and physiological signals offer a more direct view of users’ moment-to-moment experiences and are less influenced by these biases \cite{Trewin2015, Kosch2022, Frantzi2025}. Our work builds on this perspective by applying appraisal theory to ultra-short survey interactions, investigating relationships among task demands, self-experience, and psychophysiological signals.

\section{User Study}
Building on prior research, this study examines how designed survey task properties and subjective experiences jointly shape users’ physiological and behavioral responses during web-based questionnaires. Specifically, we ask:

\begin{enumerate}
\item [\textbf{RQ1:}] How do physiological and behavioral signals (EDA, HRV, eye and mouse dynamics) reflect survey-designed task demand (e.g., question difficulty, time pressure)?
\item [\textbf{RQ2:}] How do physiological and behavioral signals (EDA, HRV, eye and mouse dynamics) reflect experienced cognitive load and stress?
\item [\textbf{RQ3:}] To what extent can multimodal machine learning models detect or distinguish cognitive load and stress from physiological and behavioral signals, and how do feature-level contributions explain the experienced stress and cognitive load?
\end{enumerate}

\subsection{Study Design}
The experiment employed a $2 \times 2$ within-subjects factorial design, with two primary independent variables: \textsc{Question Difficulty} (easy vs. difficult) and \textsc{Stress Induction} (timer vs. no timer). The study protocol received ethical approval from the faculty’s ethics committee under reference number EK-MIS-2024-330-RE-d01.

\subsubsection{Independent Variables}
Two primary independent variables, \textsc{Question Difficulty} and \textsc{Stress Induction} were manipulated.

\textsc{Question Difficulty} was operationalized using validated German-language test batteries \cite{liepmann2007intelligenz, liepmann2010bowit}. Questions with high reported average accuracy rates (80\% ± 0.5\%) were classified as easy, while those with lower reported accuracy (61.5\% ± 0.5\%) were categorized as difficult, based on population test norms \cite{liepmann2007intelligenz, liepmann2010bowit}. This distinction provided a robust manipulation of cognitive load based on established performance benchmarks.

\textsc{Stress Induction} was manipulated through the presence or absence of a visible countdown timer as in previous work \cite{plonski2025much}. In the timer condition, participants were given 15 seconds to respond to each question, aiming to introduce temporal distress \cite{Vehko2019ExperiencedTP}. In the no-timer condition, participants had unlimited time to respond, reducing time-related pressure and enabling more reflective answering.

\subsubsection{Dependent Variables}
We extracted a set of physiological and behavioral variables on a per-trial basis to quantify participants' cognitive and affective responses during the task, based on previous cognitive-affective studies \cite{heimerl2022pupillometry, Heimerl2025, chiossi2023adapting, leipold2024detecting}. Physiological measures included mean tonic skin conductance level (SCL) and the average amplitude of nonspecific skin conductance responses (nsSCRs), derived from EDA signals. From the ECG, we computed mean HR and HRV as indicators of autonomic nervous system activity. To assess pupil-linked arousal, we calculated task-evoked pupil dilation by subtracting the average baseline pupil diameter (measured during fixation interval prior to each trial) from the mean pupil diameter during the active question period.

In addition to physiological data, we analyzed participants’ mouse movements to capture fine-grained indicators of cognitive load and response uncertainty. Specifically, we measured the number of directional changes along the vertical axis (y-flips), the number of hover events where the cursor remained stationary for more than 500 milliseconds, the cumulative duration of all such hover periods (total hover time), and the total distance traveled by the cursor during each trial. These metrics were derived from time-normalized mouse trajectories and were proxies for indecision, effortful processing, and task engagement. For further details on the preprocessing pipeline of each dependent variable, please refer to \autoref{sec:preprocessing}.

\subsubsection{Question Content}
To diversify the content and avoid repetition, we included the variable \textsc{Question Type}, distinguishing between general knowledge (e.g., historical events, scientific facts) and cognitive reasoning (e.g., arithmetic, sentence completion) independent of specialized domain expertise \cite{liepmann2007intelligenz, liepmann2010bowit}. While not part of our main factorial design, this variable followed the structure of the question source and was later incorporated as a random effect in the statistical modeling.

The general knowledge questions covered a wide range of topics, including fine arts, law, health, geography, literature, history, mathematics, technology, biology, and religion. These topics were counterbalanced across experimental conditions to ensure even distribution and to control for topic-specific difficulty or familiarity.

The cognitive test questions were equally divided between two types: sentence completion tasks and arithmetic reasoning problems. For the arithmetic questions, participants were asked to select the appropriate mathematical operator (e.g., +, -, ×, ÷) to correctly complete a given equation. Both sentence completion and arithmetic question types were also counterbalanced across conditions to ensure balanced exposure and minimize order or content effects. All questions were presented in German and had five answer options, including one labeled `None of the others' to minimize successful guessing through elimination strategies while reducing cognitive load \cite{dibattista2014none, elkjaer2024estimating}. Each question had only one correct answer, and participants were required to select one option before proceeding.

\subsubsection{Randomization and Counterbalancing}
\label{sec:randomization}
Participants completed a total of 48 multiple-choice questions, grouped into eight blocks of six questions each. We implemented three levels of (partial) randomization:
\begin{enumerate}
    \item \textbf{Blocks:} Each block was associated with a unique combination of the independent variables. The order of block presentation was partially randomized to control for potential sequence effects while ensuring systematic variation in the stress condition. Specifically, the blocks alternated between ``Time Stressor'' and ``No Time Stressor'' conditions to avoid the confounding effects of fatigue or habituation associated with prolonged exposure to either stress condition. We counterbalanced the stress induction condition of the first experimental block across participants.
    \item \textbf{Questions:} Within each block, the order of the six questions was fully randomized to minimize any potential order effects on performance or engagement. This randomization helped ensure that any variation in participant responses for a given question could be attributed to the experimental manipulations rather than the position of a question within the block.
    \item \textbf{Answers:} The on-screen order of each question's response options was independently randomized to prevent systematic biases or response patterns related to the placement of correct or incorrect options, thus enhancing the internal validity of the task.
\end{enumerate} 

\subsection{Apparatus}
\label{sec:apparatus}
The experimental setup integrated a combination of web-based task presentation and multimodal sensor data collection, as illustrated in \autoref{fig:overview}.

\subsubsection{Web Interface and Task Presentation} The multiple-choice questions were presented using a MacBook Pro running a custom-built Flask web application, which served as the core experimental interface. This interface was responsible for rendering questions, capturing responses, and logging time-stamped behavioral events for each interaction. All web activity, including button clicks, cursor movements, and page transitions, was continuously recorded on a local server for later synchronization with physiological signals. Before each question, participants viewed a fixation cross designed according to recommendations from \cite{thaler2013best}, combining a bullseye and crosshair to optimize both fixation stability and microsaccade suppression. All on-screen text, including question content and answer options, was rendered in a monospaced font to ensure uniform character width and maximize control over the layout of visual stimuli \cite{sharafi2020practical}.

\subsubsection{ECG Recording} We acquired ECG data at a 130\,Hz sampling rate using a Polar H10 chest strap (Polar, Finland). Before recording, the electrode was moistened with lukewarm water and placed just below the chest muscles, over the xiphoid process of the sternum, ensuring proper contact and reducing noise.

\subsubsection{EDA Recording} EDA was recorded using the BITalino (r)evolution kit (PLUX Wireless Biosignals, Portugal) at 500 Hz. Two electrodes were attached to the index and middle fingers of the non-dominant hand, after the skin was treated with Measury Potassium Chloride (KCl) Solution to enhance conductivity.

\subsubsection{Eye Tracking Recording} We collected continuous measurements of pupil diameter and blink behaviors using the Pupil Core eye-tracking headset (Pupil Labs, Berlin, Germany). The Pupil Core System consists of a dual-camera setup: (1) a scene camera positioned forward to capture the wearer’s field of view in high-definition (HD) video at 30 Hz, and (2) an infrared eye camera for each eye that records monocular eye movements at a high sampling rate of 120 Hz. The scene camera provided a first-person perspective of the participant’s visual environment, while the infrared eye cameras tracked gaze position and pupil dynamics. All eye-tracking data, including pupil diameters and blink events, were processed and exported using Pupil Player, the proprietary software developed by Pupil Labs. 

\subsubsection{Mouse Tracking Recording} Participants interacted with the web interface using a Logitech mouse, operated with their dominant hand. During the entire session, the position of the mouse cursor on screen (x and y coordinates) and corresponding timestamps were recorded through mousetrap-web \cite{Wulff2025}. This allowed for reconstruction of motor behavior and cursor trajectories during decision-making processes.

\subsubsection{Stimuli Synchronization} Sensor data streams were acquired and temporally synchronized using the Lab Streaming Layer \cite{kothe2024lab}, which assigns high-resolution timestamps at acquisition and continuously synchronizes clocks across devices to correct for drift, recorded using LabRecorder. Additional synchronization was performed using the timestamped logs from the Flask web application, allowing precise alignment between signal recordings and specific task events (e.g., stimulus onset, response time, rating screens).

\subsection{Procedure}
\label{sec:procedure}
Upon arrival, participants were seated at a table and received a detailed verbal and written overview of the study, including its goals, task structure, data collection methods, and equipment. They were informed that eye movements, mouse interactions, and physiological signals would be continuously recorded. After providing informed consent, participants completed a short demographic questionnaire covering age, gender, handedness, and German language proficiency.

Participants were then fitted with the sensors described in \autoref{sec:apparatus}. To establish physiological baselines, a three-minute resting period followed, during which participants sat still and relaxed.
Then participants completed the eight blocks of the experimental session, see \autoref{sec:randomization}.

\begin{figure}[htbp]
    \centering
    \includegraphics[width=1.0\textwidth]{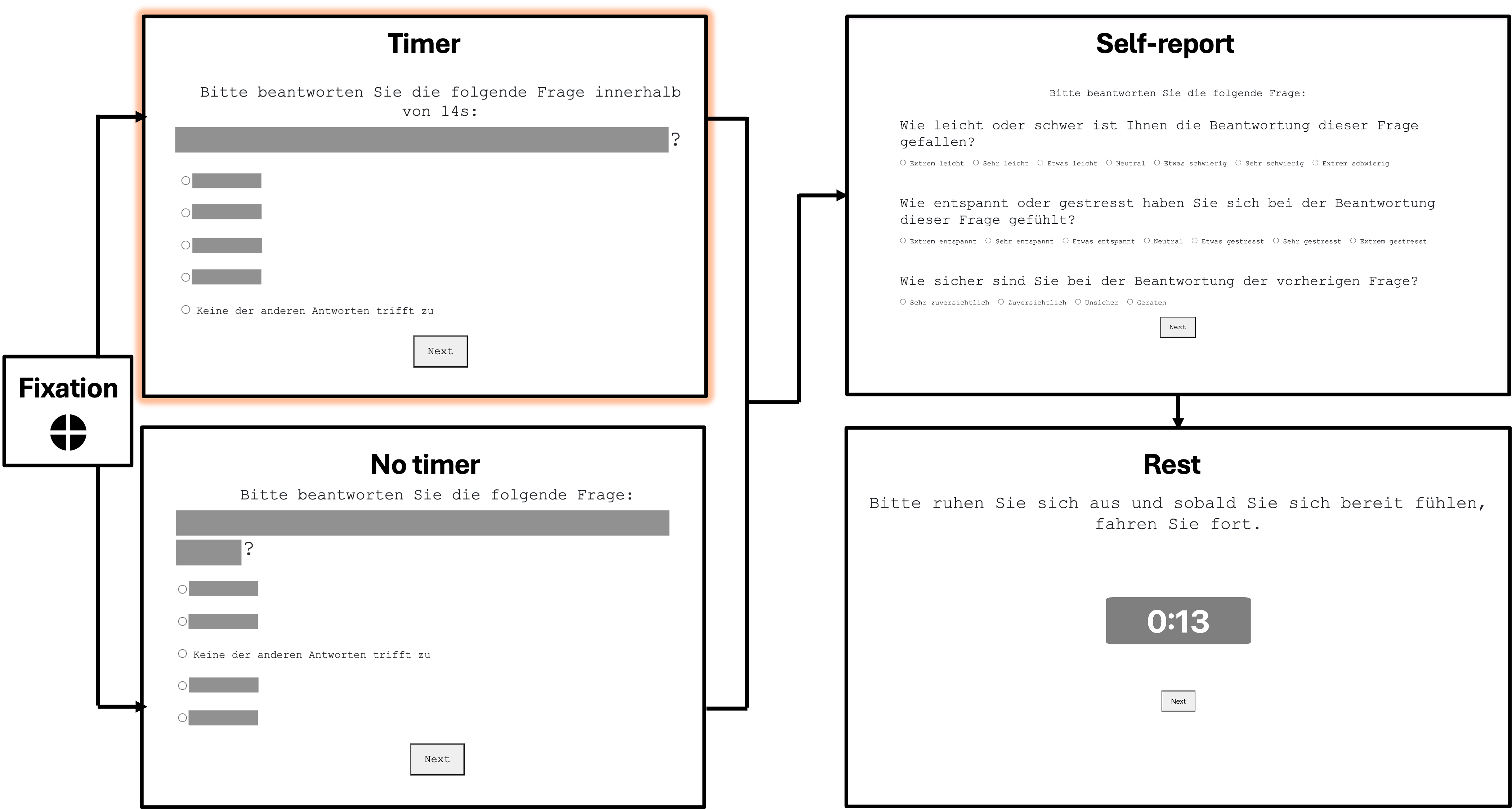} 
    \caption{Flow diagram illustrating the procedure of trials (questions). (1) fixation cross, (2) Multiple-choice question with timer or no timer, (3) Self-report on question difficulty, subjective stress, and confidence in the provided answer, (4) Suggested 20-second rest period with countdown}
    \Description{Flow diagram illustrating the procedure of trials (questions). (1) fixation cross, (2) Multiple-choice question with timer or no timer, (3) Self-report on question difficulty, subjective stress, and confidence in the provided answer, (4) Suggested 20-second rest period with countdown}
    \label{fig:page}
\end{figure}

As shown in \autoref{fig:page}, each trial began with a fixation cross displayed at the center of the screen for a randomized duration of 1250 ms, 1500 ms, or 1750 ms, before disappearing automatically \cite{chiossi2024understanding}. This fixation period was designed to standardize visual attention and minimize carryover effects, following best practice recommendations \cite{thaler2013best}. Following the fixation, the multiple-choice question appeared on screen. After selecting and confirming their answer, a 1-second inter-stimulus interval was presented without any cross or objects to allow for neural and attentional reset, counteracting fatigue effects \cite{atkinson1989and, woodman1999electrophysiological}. Once an answer was submitted by clicking the “Next” button, participants could not return to or revise their previous responses. Then, participants were prompted to complete three rating scales assessing their subjective experience:

\begin{itemize}
    \item \textsc{Perceived difficulty} – “How easy or difficult was it for you to answer this question?” (7-point scale
from 1 [extremely easy] to 7 [extremely difficult])

    \item \textsc{Perceived stress} – “How relaxed or stressed did you feel when answering this question?” (7-point scale from 1 [extremely relaxed] to 7 [extremely stressed])

    \item \textsc{Confidence} – “How confident are you in your answer to the previous question?” (4-point scale from 1 [very confident] to 4 [guessed]) \cite{kennel2025gaze}
\end{itemize}

Following these ratings, the interface suggested a resting period of 20 seconds, but participants were free to skip the resting period or stay longer than the 20 seconds before clicking the “Next” button to proceed with the next question. A five-minute break was suggested halfway through the session to minimize cognitive fatigue.

After task completion, sensors were removed and sanitized, and participants were offered monetary compensation. 

\subsection{Participants}
A total of 30 participants were recruited via advertisements circulated throughout the university campus. All experimental sessions were conducted in a controlled laboratory environment with standardized lighting, noise levels, and workstation setup to ensure consistent testing conditions. Data from one participant were excluded from the final analysis because all self-reported stress ratings were neutral, resulting in zero variance across assessments, suggesting potential non-compliance or disengagement with the subjective assessments. Therefore, the final dataset comprised 29 valid participants. 
Participants ranged in age from 19 to 74 years (Mean = 27.03, SD = 11.07). The sample included 12 male, 17 female participants, none diverse. Twenty participants were native German speakers, 8 reported an advanced level of German proficiency, and 1 participant indicated an intermediate level of proficiency. All participants self-identified as right-handed. They were compensated with either 10 euros or course credits. 

\section{Data Analysis}

\subsection{Pre-processing and Feature Extraction}
\label{sec:preprocessing}
We preprocessed EDA signals using NeuroKit2 \cite{makowski2021neurokit2}. The raw EDA data were first filtered with a third-order Butterworth high-pass filter (3 Hz) to remove slow drifts and low-frequency artifacts. Following this, we applied nonnegative deconvolution analysis \cite{benedek2010decomposition} to decompose the signal into tonic (skin conductance level, SCL) and phasic (nonspecific skin conductance responses, nsSCRs) components. To quantify EDA activity, we computed mean tonic SCL as a baseline measure of sympathetic arousal and average amplitude of nsSCRs, with nsSCR peaks detected using a threshold of 0.05 $\mu$S to exclude minor fluctuations. ECG signals were analyzed in the time domain to derive HR and HRV metrics.
The raw ECG data were band-pass filtered (3–45 Hz, 3rd-order FIR filter) to remove motion artifacts and high-frequency noise. R-peak detection was performed using Hamilton’s segmentation method \cite{hamilton2002open}, followed by correction of ectopic beats and artifacts. We computed mean HR as a global measure of autonomic activity and HRV to assess parasympathetic modulation.
For eye movement analysis, pupil diameters and blink counts are extracted from the eye with higher tracking confidence (using Pupil Core's Python API during initial data cleaning). Data points with confidence values below 50\% were excluded to ensure signal quality. Additionally, physiologically implausible pupil diameter values (>50 mm, likely due to tracking errors) were removed as outliers. To assess task-evoked pupillary responses, for each trial, we calculated the pupil difference as mean pupil diameter during the task period minus mean pupil diameter during fixation baseline. This metric represents the absolute change in mean pupil size between the pre-stimulus baseline and active task engagement periods.
For each question, we analyzed participants’ mouse movements by extracting four trajectory-based indices relevant to our hypotheses: the number of y-flips, number of hovers, total hover time, and total distance traveled. Prior to feature extraction, all mouse trajectories were time-normalized using the interpolation functions provided by the mousetrap package \cite{kieslich2019mouse}. This process resampled each trajectory into a fixed number of equally spaced time intervals. The number of hovers was computed by identifying how often the mouse cursor remained stationary for longer than 500 milliseconds, excluding the initial latency period before the participant began moving the mouse at the start of each trial, providing an possible index of movement hesitation or indecision. Total hover time was defined as the cumulative duration of all such stationary periods within a trial, which could reflect overall hesitation during response selection. The number of y-flips was calculated by counting the number of directional reversals along the vertical (y) axis. Finally, total distance was computed as the sum of Euclidean distances between successive (x, y) coordinates in the time-normalized trajectory, representing the overall extent of cursor movement.

\begin{figure}[htbp]
    \centering
    \begin{subfigure}[b]{0.48\textwidth}
        \includegraphics[width=\textwidth]{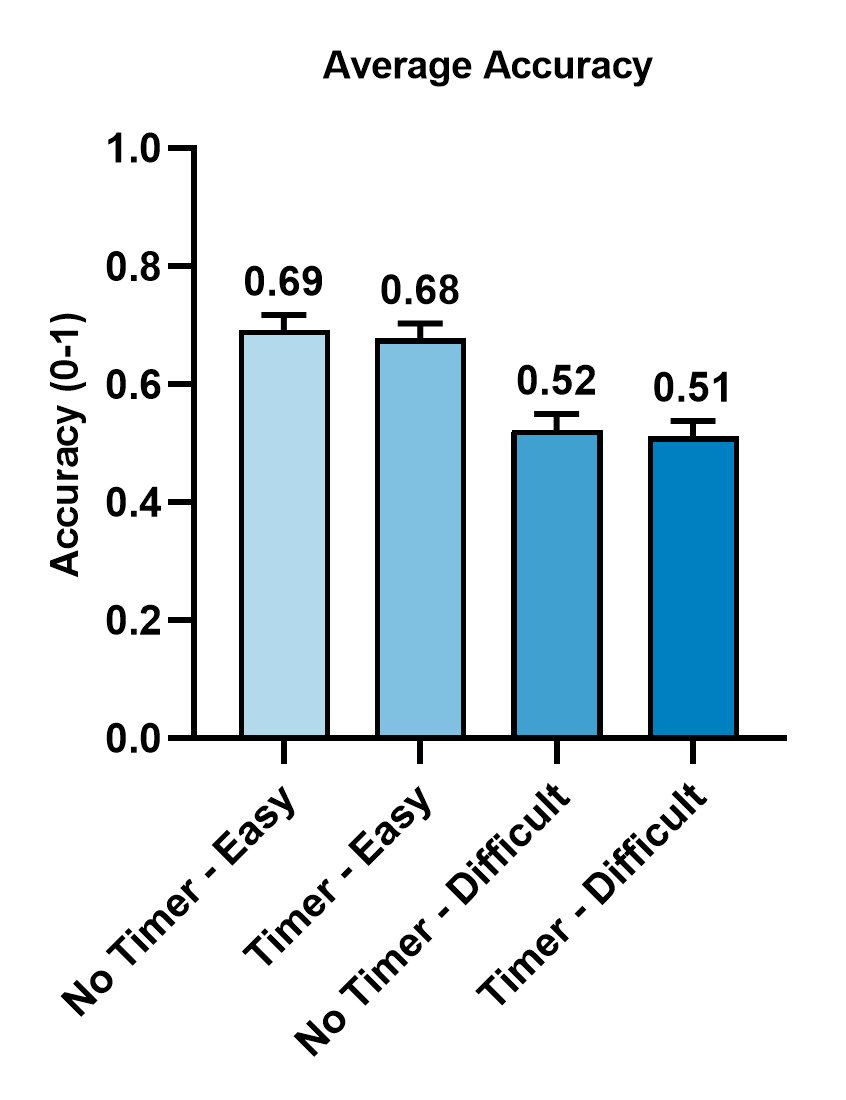}
        \caption{Mean accuracy over manipulated conditions with standard error}
        \label{fig:acc}
    \end{subfigure}
    \hfill
    \begin{subfigure}[b]{0.48\textwidth}
        \includegraphics[width=\textwidth]{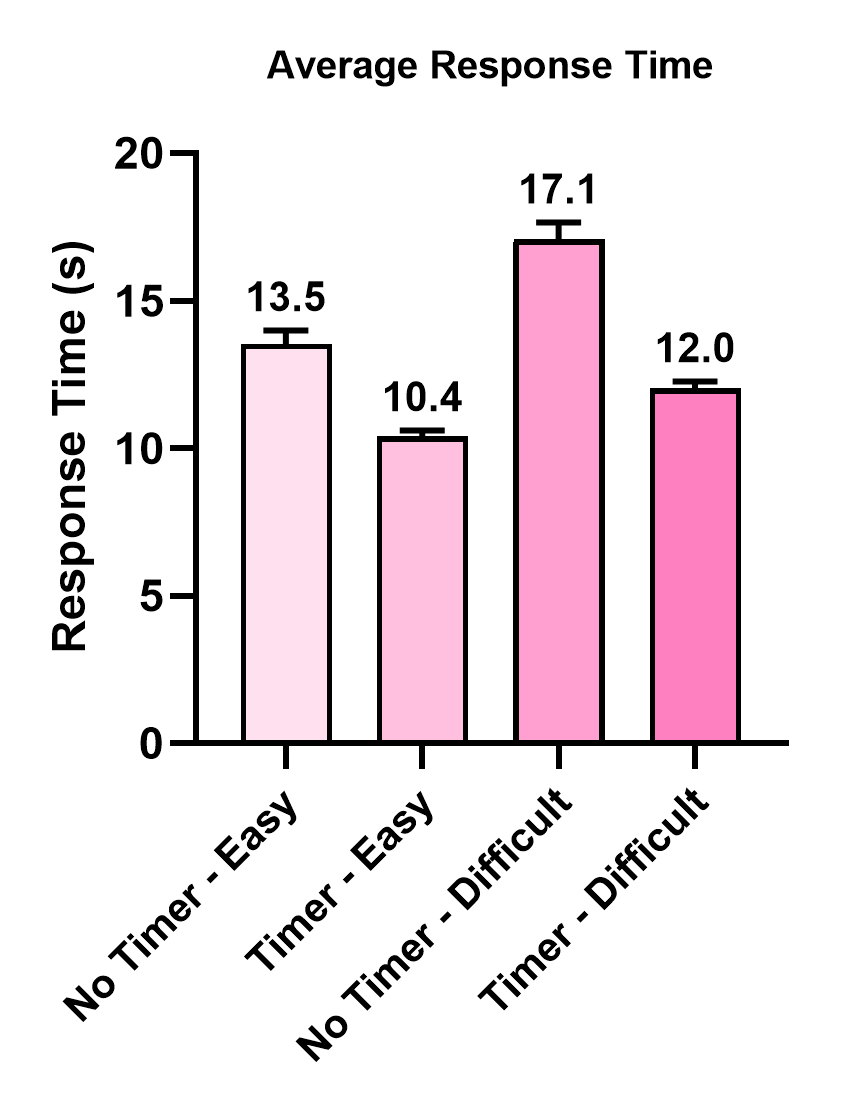}
        \caption{Average response time over manipulated conditions with standard error}
        \label{fig:rt}
    \end{subfigure}
    
    \centering
    \caption{The (a) average accuracy and (b) the average response time across the four experimental conditions with standard errors formed by the interaction of question difficulty and time pressure. Accuracy was higher for easy tasks (0.69 without timer, 0.68 with timer) compared to difficult tasks (0.52 without timer, 0.51 with timer). Response times were consistently longer without time pressure, indicating that both question difficulty and time constraints impacted response behavior. }
    \label{fig:bar}.
\end{figure}

\subsection{Data Labeling and Manipulation Check}
As described in \autoref{sec:procedure}, our experiment varied question difficulty (easy vs. difficult) and the presence or absence of time pressure (timer vs. no timer) as measures of participants' stress and cognitive load levels. Accordingly, we assigned categorical labels to every trial. Furthermore, our second labeling approach relied on participants' subjective self-reports, recording their perceived stress and perceived difficulty on 7-point scales following each question. For these ratings, descriptive statistics and visualization are shown in \autoref{sec:sdp} and \autoref{sec:vis}.

To assess whether the experimental manipulations were effective, we conducted Wilcoxon signed-rank tests to evaluate whether participants’ self-reported stress and difficulty ratings differed depending on the assigned experimental conditions. Indeed, the differences were statistically significant indicating differences in perceived cognitive and affective states between the manipulation settings. In detail, the test assessing differences in the self-reported difficulty between the easy and difficult conditions yielded a Wilcoxon statistic of 10.0 with a $p$-value of \num{1.68e-5}, and the test for the time-stress manipulation (with vs. without timer) resulted in a Wilcoxon statistic of 40.0 and a $p$-value of 0.0002.

As shown in \autoref{fig:bar}, the average accuracy rates for questions of the same difficulty level remained consistent under different timer conditions, suggesting that time pressure did not significantly alter participants' ability to answer correctly when difficulty was held constant. Further, the average response times across different conditions revealed consistent patterns aligning with the question difficulty and stress manipulations. Under conditions without the timer, participants took longer to respond on average, especially for difficult questions. Specifically, for difficult questions without time pressure, the mean response time was 17.079 seconds (SD = 10.851), compared to 12.029 seconds (SD = 4.372) for difficult questions with a timer. For easy questions, the mean response time without a timer was 13.545 seconds (SD = 8.550), while with a timer it decreased to 10.400 seconds (SD = 3.965). These findings further support the effectiveness of the experimental design in manipulating both question difficulty and time-induced stress, as well as their measurable impact on participant behavior.

\subsection{Investigating Relationships among Design, Experience, and Psychophysiological Signals}
The physiological and behavioral data collected during the study were analyzed at the level of individual questions to capture fine-grained variations in participant responses. Each question block was classified into binary given stress (timer vs. no timer) and difficulty (easy vs. difficult) conditions based on the experimental manipulations. These binary conditions served as fixed effects in the statistical analysis. To assess the effects of these manipulations on participants’ physiological and behavioral responses, we employed mixed-effects linear models. These models included the physiological and behavioral metrics as dependent variables, while accounting for stress and difficulty as fixed predictors. To account for additional sources of variability without overfitting the model, we treated certain factors as random effects. Individual participants were modeled with random intercepts and slopes for trial order to capture both baseline differences and individual-specific trends across the task sequence, such as learning, fatigue, or habituation effects. Question type was modeled as a random variable to account for variability associated with different question categories. In parallel, we analyzed self-reported stress and difficulty, which were collected after each question on 7-point scales. To align the analysis with the semantic structure of the scales, where 1 (“extremely relaxed/easy”) to 3 represent clearly low states, 4 reflects a conceptually neutral midpoint, and 5 to 7 (“extremely stressed/difficult”) indicate increasingly high states, we recoded the responses into three ordered categories: Low (1–3), Neutral (4), and High (5–7). Collapsing categories in this way reduces spurious variability arising from distinctions that participants may not reliably make, preserves the ordinal meaning of the scale, and supports more interpretable comparisons across qualitatively distinct levels of self-reported stress and difficulty. The subjective ratings were then analyzed using a linear mixed-effects model, predicting performance from the ordered categorical stress and difficulty factors and their interaction. The model included random intercepts and slopes for participant-to-participant order, and random intercepts of question type as well. 

\subsection{Investigating Feature Importance for Detecting Cognitive Load and Stress}

\subsubsection{Modeling Approach and Feature Extraction}
To investigate the extent to which multimodal machine learning can detect and distinguish cognitive load and stress from physiological and behavioral signals, we developed and evaluated a series of classifiers. Our modeling approach proceeded in two stages: we first trained models to predict the binary, experimentally manipulated conditions (i.e., presence of a timer for stress and manipulated question difficulty for cognitive load) to establish a baseline for what is detectable from the signals; we then trained models to predict participants' self-reported cognitive load (difficulty) and stress on a per-question basis, providing a direct measure of the experienced states. For self-reports, neutral responses (ratings of 4) were excluded to maintain a clear distinction between positive and negative states. Including them would have increased class imbalance and potentially obscured the patterns the models were designed to detect \cite{Zou2016}. In both cases, combining the binary stress and difficulty labels resulted in four distinct classes reflecting cross-conditions: ``relaxed \& easy'', ``relaxed \& difficult'', ``stressed \& easy'', and ``stressed \& difficult''.

We extracted a comprehensive set of features from all available data modalities, including physiological signals, behavioral measures, and response times (cf. \autoref{sec:preprocessing}). To control for inter-individual differences in baseline physiology and behavior, all features were normalized via per-subject z-normalization prior to model training \cite{park2024hide, 10.1145/3432220}.

\subsubsection{Classifier Selection and Validation Strategies}
We evaluated three distinct classifiers---Support Vector Machine (SVM), k-Nearest Neighbors (KNN), and Random Forest (RF)---to ensure our findings were not specific to a single algorithm. To evaluate model performance and generalizability while reducing overfitting, we employed 10-fold cross validation and Leave-one-subject-out (LOSO) cross-validation. The first approach, 10-fold cross-validation, assessed the model's ability to generalize to unseen data from the same participants. To prevent data leakage and over-optimistic performance estimates, hyperparameter grid search was conducted within the training folds of each cross-validation split. The second approach, Leave-One-Subject-Out cross-validation, provided a more stringent test of the model's ability to generalize to completely new, unseen participants, which represents the central challenge in affective computing. In LOSO, a model is trained on data from all but one participant and tested on the held-out participant; this process is repeated iteratively until every participant has served as the test set. This approach inherently prevents participant-level overfitting and provides a realistic estimate of real-world performance.

For both validation strategies, hyperparameters were tuned using grid search executed within the training portion of each 3-fold split (nested cross-validation) for each configuration. The outer test folds were not used for tuning and remained strictly reserved for unbiased performance evaluation. Model performance was assessed using both accuracy and macro F1-score, with the latter providing a robust measure in the presence of potential class imbalance.

\subsubsection{Feature Importance and Interpretability Analysis}
To answer the question of how feature-level contributions explain the experienced states, we computed SHapley Additive exPlanations (SHAP) values \cite{10.5555/3295222.3295230} for the best-performing model. SHAP analysis provided global interpretability, revealing the overall contribution of each feature and data modality to the model's predictions for cognitive load and stress.

\section{Results}
All model training and evaluation experiments were performed on a Macbook Pro M3. Data handling and model development were implemented in Python. The linear mixed models from the Pymer4 \cite{jolly2018pymer} wrapper for the lme4 package \cite{bates2015fitting} were used to test the effect of our manipulations or self-reported ratings on the dependent measures. Machine learning model training was implemented using the scikit-learn library.

\subsection{Associations Between Task Demands and Psychophysiological Signals}
We modeled the effects of experimental manipulations on the behavioral and physiological dependent variables (i.e., the features described in \autoref{sec:preprocessing}) using the formula: dependent variable $ \sim \texttt{stress induction} \times \texttt{question difficulty} + (\texttt{question order}|\texttt{participant id}) + (1|\texttt{question type})$. The interaction term between stress induction and question difficulty was included, rather than modeling them only additively, because our study design was a fully crossed 2×2 within-subjects setup, where all combinations of conditions (timer/no timer × easy/dfficult) were presented in a counterbalanced order. Our results are shown in \autoref{tab:model_results_c}. 
Increased task difficulty was associated with a higher number of mouse flips on the y-axis, more frequent hovers, longer hover time, greater total mouse distance, and increased blink count and duration. In contrast, time-stress was linked to a significant decrease in hovers, hover time, blink count, and blink duration, alongside an increase in tonic EDA and total mouse distance. SCR peaks showed a divergent pattern: time-stress decreased the number of peaks, while difficulty increased it. The interaction between difficulty and time-stress was not significant for any of the dependent measures.

\begin{table}[htbp]
\centering
\caption{Mixed linear model results, using the experimental manipulations as independent variables}
\label{tab:model_results_c}
\begin{tabular}{l
                S[table-format=1.3]@{\hspace{12pt}}l
                S[table-format=1.3]@{\hspace{12pt}}l
                S[table-format=1.3]@{\hspace{12pt}}l
                S[table-format=1.3]@{\hspace{12pt}}l}
\toprule
\textbf{Dependent Variable} & \multicolumn{2}{c}{\textbf{Timer}} & \multicolumn{2}{c}{\textbf{Difficult}} & \multicolumn{2}{c}{\textbf{Interaction}} & \multicolumn{2}{c}{\textbf{Intercept}}\\
\cmidrule(r){2-3} \cmidrule(r){4-5} \cmidrule(r){6-7}\cmidrule(r){8-9}
 & {Est.} & {$p$} & {Est.} & {$p$} & {Est.} & {$p$}& {Est.} & {$p$} \\
\midrule
Mouse y flips & 0.018&0.937&\textbf{0.352}&\textbf{0.001}&0.008&0.930&5.862&<0.001\\

Hovers&\textbf{-0.257}&\textbf{<0.001}&\textbf{0.118}&\textbf{0.004}&0.019&0.622&1.737&<0.001\\

Hover time&\textbf{-835.770}&\textbf{<0.001}&\textbf{294.736}&\textbf{0.009}&\text{-50.502}&0.621&\text{3232.9}&0.044\\

Total distance&\textbf{34.978}&\textbf{0.002}&\textbf{41.129}&\textbf{0.002}&1.878&0.868&\text{1157.4}&<0.001\\
\hline
Pupil difference&\text{-0.009}&0.562&\text{0.001}&0.959&0.003&0.827&0.671&<0.001\\

Blink count&\textbf{-0.393}&\textbf{<0.001}&\textbf{0.174}&\textbf{0.013}&\text{-0.050}&0.437&1.171&0.419\\

Blink duration&\textbf{-0.087}&\textbf{<0.001}&\textbf{0.037}&\textbf{0.014}&\text{-0.010}&0.455&0.251&0.445\\
\hline
EDA tonic average&\textbf{11.420}&\textbf{<0.001}&-2.902&0.245&3.851& 0.053&\text{390.116}&<0.001\\

EDA amplitude&0.001&0.471&0.001&0.428&\text{-0.001}&0.841&0.006&0.213\\

SCR peaks&\textbf{-0.981} & \textbf{<0.001}&\textbf{0.658}&\textbf{0.022}&0.366&0.135&10.266&<0.001\\
\hline
HR&\text{-0.186}&0.141&\text{-0.086}&0.558&0.203&0.101&79.991&<0.001\\

HRV&\text{-0.688}&0.877&3.834&0.419&\text{-3.067}&0.485&104.952&0.413\\
\bottomrule
\end{tabular}
\end{table}

\subsection{Associations Between Self-Reports and Psychophysiological Signals}
Based on the mixed linear model results using self-reported stress and difficulty ratings as independent variables, shown in \autoref{tab:model_results_self}, several physiological and behavioral measures showed significant associations with perceived cognitive load and stress. Self-reported difficulty was a significant predictor of increased mouse flips, hovers, hover time, total cursor distance, blink count, blink duration, and SCR peaks. In contrast, self-reported stress was a significant predictor of increased mouse flips, pupil response, and EDA tonic average, while also predicting a decrease in EDA tonic average. A significant interaction between stress and difficulty was found for hovers, indicating that the combined experience of stress and difficulty had a unique effect on this specific signal.

\begin{table}[htbp]
\centering
\caption{Mixed linear model results for taking the Self-reported Ratings as independent variables}
\label{tab:model_results_self}
\begin{tabular}{l
                S[table-format=1.3]@{\hspace{12pt}}l
                S[table-format=1.3]@{\hspace{12pt}}l
                S[table-format=1.3]@{\hspace{12pt}}l
                S[table-format=1.3]@{\hspace{12pt}}l}
\toprule
\textbf{Dependent Variable} & \multicolumn{2}{c}{\textbf{Stress Rating}} & \multicolumn{2}{c}{\textbf{Difficult Rating}} & \multicolumn{2}{c}{\textbf{Interaction}} & \multicolumn{2}{c}{\textbf{Intercept}}\\
\cmidrule(r){2-3} \cmidrule(r){4-5} \cmidrule(r){6-7}\cmidrule(r){8-9}
 & {Est.} & {$p$} & {Est.} & {$p$} & {Est.} & {$p$}& {Est.} & {$p$} \\
\midrule
Mouse y flips & \textbf{0.314} & \textbf{0.027} & \textbf{0.493} & \textbf{<0.001}&0.233&0.095&6.096&\text{<0.001}\\

Hovers&\text{-0.024}&0.683&\textbf{0.308}&\textbf{<0.001}&\textbf{-0.167}&\textbf{0.004} & \text{1.807}&\text{<0.001}\\

Hover time&280.502&0.081&\textbf{588.204}&\textbf{<0.001}&\text{-178.741}&0.266 & \text{3385.3}&\text{<0.001}\\

Total distance&29.018&0.106&\textbf{64.647}&\textbf{<0.001}&\text{20.125}&0.256&\text{1152.7}&\text{<0.001}\\
\hline
Pupil difference&\textbf{0.070}&\textbf{0.003}&\text{-0.004}&0.840&\text{-0.029}&0.212&0.694&\text{<0.001}\\

Blink count&\text{-0.037}&0.715&\textbf{0.215}&\textbf{0.021}& \text{-0.118}&0.246&1.274&0.176\\

Blink duration&\text{-0.008}&0.713&\textbf{0.047}&\textbf{0.019}&\text{-0.023}&0.301&0.269&0.449\\
\hline
EDA tonic average&\textbf{13.186}&\textbf{<0.001}&\textbf{-13.304}&\textbf{<0.001}&\text{2.484}&0.438&\text{391.03}&\text{<0.001}\\

EDA amplitude&\text{-0.001}&0.752&\text{-0.001}&0.981&0.001&0.329&0.006&0.327\\

SCR peaks&\text{-0.013}&0.973&\textbf{1.974}&\textbf{<0.001}&\text{-0.239}&0.536&\text{10.367}&\text{<0.001}\\
\hline
HR&\text{-0.098}&0.619&\text{-0.288}&0.107&\text{-0.121}&0.529&\text{79.653}&\text{<0.001}\\

HRV&\text{0.281}&0.967&\text{-2.290}&0.717&\text{-6.836}&0.313&\text{113.156}&0.405\\
\bottomrule
\end{tabular}
\end{table}

\subsection{Classification with Machine Learning and Feature Importance}

To answer RQ3, to what extent multimodal machine learning models can detect cognitive load and stress from physiological and behavioral signals, we evaluated three models (KNN, SVM, and RF) across two labeling strategies (Given Condition and Self-Reported) using 10 repeated runs per configuration with different randomization seeds, shown in \autoref{tab:merged_tasks}. For the 4-class stress × difficulty prediction, distinguishing the categories  ``relaxed \& easy'', ``relaxed \& difficult'', ``stressed \& easy'', and ``stressed \& difficult'', using self-reported labels, which most directly reflect subjective user state, model performance differed substantially. The RF model achieved the highest mean accuracy (0.52), followed by SVM (0.49) and KNN (0.39). To account for multiple comparisons, we applied a Bonferroni correction ($\alpha$ = 0.05/3 = 0.0167) to pairwise paired t-tests. All performance differences remained statistically significant (all corrected p < 0.001). Specifically, RF significantly outperformed SVM (mean difference = 0.028, t = -10.42, corrected p < 0.001) and KNN (mean difference = 0.132, t = -29.31, corrected p < 0.001), while SVM also significantly outperformed KNN (mean difference = 0.104, t = -26.10, corrected p < 0.001). Consequently, we selected the RF model with self-reported labels for subsequent SHAP analysis, as it demonstrated the most robust performance for interpreting the learned relationships between multimodal signals and the target cognitive-affective states.

\begin{table}[htbp]
\centering
\caption{Comparison of model performance across evaluation methods and labeling strategies. The right column indicates the improvement over random classification.}
\label{tab:merged_tasks}
\begin{tabular}{lllSSS}
\toprule
\multirow{2}{*}{Eval.} & \multirow{2}{*}{Model} & \multirow{2}{*}{Labeling} & \multicolumn{3}{c}{Stress × Difficulty (4 Categories)} \\
\cmidrule(lr){4-6}
& & & {Acc. (SD)} & {Macro F1 (SD)} & {Imp. over Random} \\
\midrule

\multirow{6}{*}{10-Fold} 
& \cellcolor{blue!20}KNN & \cellcolor{blue!10}Given Cond.  &\text{0.35 (0.009)} & \text{0.32 (0.009)} & \text{+0.10} \\
&  \cellcolor{blue!20}   & \cellcolor{green!10}Self-reported  & \text{0.39 (0.008)} & \text{0.34 (0.006)}& \text{+0.14} \\

& \cellcolor{green!20}SVM & \cellcolor{blue!10}Given Cond.   & \text{0.40 (0.006)} & \text{0.37 (0.007)}& \text{+0.15} \\
&   \cellcolor{green!20}  & \cellcolor{green!10}Self-reported   & \text{0.49 (0.001)} & \text{0.38 (0.002)} & \text{+0.24} \\
& \cellcolor{orange!20}RF  & \cellcolor{blue!10}Given Cond.    & \text{0.45 (0.008)} & \text{0.35 (0.007)} & \text{+0.20} \\
&  \cellcolor{orange!20}   & \cellcolor{green!10}Self-reported   & \text{0.52 (0.005)} & \text{0.35 (0.006)} & \text{+0.27} \\
\midrule

\multirow{6}{*}{LOSO} 
& \cellcolor{blue!20}KNN & \cellcolor{blue!10}Given Cond. & \text{0.32 (0.012)} & \text{0.30 (0.011)} & \text{+0.07} \\
&  \cellcolor{blue!20}   & \cellcolor{green!10}Self-reported  & \text{0.35 (0.009)} & \text{0.28 (0.009)} & \text{+0.10} \\
& \cellcolor{green!20}SVM & \cellcolor{blue!10}Given Cond.  & \text{0.35 (0.009)} & \text{0.34 (0.008)} & \text{+0.10} \\
&   \cellcolor{green!20}  & \cellcolor{green!10}Self-reported   & \text{0.48 (0.006)} & \text{0.31 (0.007)} & \text{+0.23} \\
& \cellcolor{orange!20}RF  & \cellcolor{blue!10}Given Cond.   & \text{0.35 (0.011)} & \text{0.34 (0.011)} & \text{+0.10} \\
&  \cellcolor{orange!20}   & \cellcolor{green!10}Self-reported    & \text{0.46 (0.007)} & \text{0.32 (0.009)} & \text{+0.21} \\
\bottomrule
\end{tabular}
\end{table}

\begin{figure}[htbp]
\includegraphics[width=\textwidth]{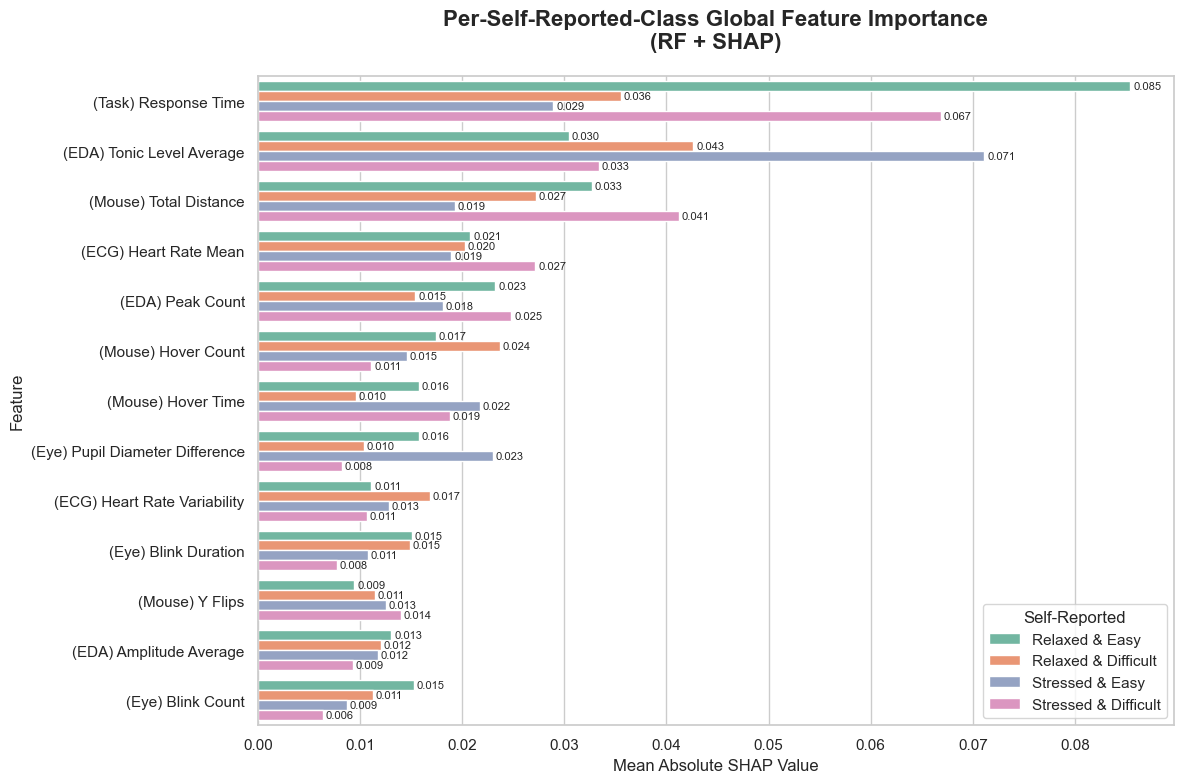}
        \caption{SHapley Additive ex-Planations (SHAP) for all features used in the RF classifier, including 4-self-reported-class labels}
        \label{fig:shap}
\end{figure}

Our SHAP analysis (\autoref{fig:shap}) reflects how different sensor modalities contribute to classifying users into the four self-reported cognitive-affective states. These results, based on the RF classifier, show patterns across behavioral and physiological features that align with theoretical expectations of cognitive load and emotional stress. Notably, response time emerged as the most influential feature in distinguishing between ``relaxed \& easy'' and ``stressed \& difficult'' states. This results shows the impact of task fluency and decisiveness as key behavioral indicators of low versus high cognitive-affective burden. However, when trained using response time alone, a random forest classifier with 10-fold cross-validation achieved an average accuracy of only 0.2706 (±0.0764), substantially below the performance of the full multimodal model. This suggests that while response time is important, it is insufficient on its own to reliably capture the nuanced differences across all four states. In addition, some features, i.e., counts or durations, are likely to be correlated with the response time. Tonic EDA average was the most important feature for classifying the two mixed states, ``relaxed \& difficult'' and ``stressed \& easy''. 
This physiological measure, reflecting sympathetic arousal over sustained periods, appears to capture a latent dimension of physiological activation that is not directly mirrored in task performance. The complementary nature of EDA and response time is particularly noteworthy: while response time reflects task-related effort or fluency, tonic EDA provides a window into the user’s internal arousal level, even when the behavioral output is ambiguous. 
For example, someone might complete an easy task quickly but still feel stressed, a subtlety that behavioral data alone may miss. Beyond these dominant features, we also observed that mouse trajectory distance notably contributed to classifying the ``stressed \& difficult'' state.

\section{Discussion}
In this work, we evaluated the relationship among induced time-stress manipulation and question difficulty,  behavioral and physiological data, and perceived cognitive-affective status.

\subsection{Summary of Results}

\subsubsection{When Design and Experience Align}
Across both objective manipulations and self-reported experience, several behavioral and physiological patterns converged, revealing conditions under which task design and lived experience carry consistent information. Higher task difficulty reliably produced markers of increased cognitive effort, more mouse flips and hovers, longer hover time, greater cursor distance, more SCR peaks, and more frequent and longer blinks, and these same hesitation and effort signatures also emerged for self-reported difficulty. This parallel suggests that both externally imposed demands and internally appraised demands shape micro-level interaction behaviors in somewhat similar ways. Likewise, both time-pressure and self-reported stress were associated with elevated tonic EDA, indicating that heightened arousal manifests consistently across objective constraints and subjective stress interpretations \cite{Awada2024StressAI}. These convergences highlight the circumstances under which designers’ expectations about difficulty or stress map cleanly onto the multimodal signals that adaptive systems can leverage. For the user-centered design of surveys and similar interfaces, this alignment is important: it suggests that certain behavioral and physiological markers, particularly fine-grained cursor hesitation patterns and blink dynamics, are robust indicators of cognitive effort regardless of whether that effort stems from the task as designed or from how the respondent experiences it. In these cases, task characteristics and subjective experience are reflected through a coherent pattern, offering a stable foundation for real-time adaptive survey systems to detect and interpret respondent state.

\subsubsection{When Design and Experience Misalign}
However, the results also reveal systematic divergences between objective manipulations and subjective experience, confirming appraisal theory that task demands doesn't always equal to perceived experience \cite{Lazarus1991}. Time-stress, for example, produced a distinct behavioral compression, reducing hovers, blink activity, and hover time, while also increasing total mouse distance and decreasing SCR peaks. Yet self-reported stress did not reproduce this pattern. Instead, it was associated with more mouse flips, larger pupil differences, and greater tonic EDA, indicating a more internally oriented arousal response rather than the outward behavioral streamlining elicited by imposed time pressure. Additional dissociations emerged as well: decreases in tonic EDA were significant only in self-reported difficulty, whereas reductions in SCR peaks were detectable only under experimentally induced time-stress. Further, the only significant interaction between self-reported stress and difficulty occurred for hovers, indicating that the combined experience of stress and difficulty produces a unique behavioral signature not predictable from either factor alone. These divergences highlight the complexity of the subjective stress experience. Participants’ reported stress levels might not map cleanly onto the time-pressure manipulation; instead, stress appeared to arise from cognitive appraisals that integrated both time constraints and perceived difficulty. This aligns with theoretical accounts positioning stress not as a direct consequence of a single environmental factor, but as an emergent evaluation of task demands relative to personal resources \cite{sandi2013stress, al2023essential}. Participants occasionally reported high stress on objectively “easy” questions when time pressure was high, and low stress on “difficult” questions when time pressure was absent, illustrating how individual differences in appraisal shape subjective experience in ways that depart from designers’ expectations \cite{Chyu2022AssociationsBA}.
These discrepancies underscore a critical insight for HCI: objective task manipulations and subjective experience can generate qualitatively different multimodal signatures, even when intended to target the same construct. For designers of adaptive survey systems, this means that real-time sensing must accommodate the possibility that respondents reinterpret or reweigh task demands according to their own capacities and strategies. Misalignment is therefore not noise. It is meaningful information about how individuals negotiate overlapping demands and internal states. Understanding when and why these divergences arise is essential for developing adaptive interfaces that respond to experienced, rather than merely assumed, difficulty or stress.

\subsubsection{Insights from Multimodal Models and SHAP Analyses}
The machine-learning results offer a complementary perspective on how cognitive load and stress manifest across multimodal signals. Consistent with our behavioral and physiological analyses, models trained on self-reported states substantially outperformed those trained on objective task conditions, reinforcing the importance of subjective appraisal in shaping measurable user states. Among the three classifiers, the RF model achieved the highest accuracy on the four self-reported cognitive–affective categories, with performance substantially exceeding both SVM and KNN. This superior performance suggests that cognitive–affective states during brief survey interactions are best captured by nonlinear interactions across modalities, aligning with theories that cognitive load and stress arise from coupled behavioral and physiological processes rather than independent signals \cite{Matthews2014, Calvo2010}. SHAP analysis further illuminates how these multimodal features contribute to classification and provides insight into the underlying cognitive–affective mechanisms. Response time emerged as the most influential feature, strongly separating the “relaxed \& easy” and “stressed \& difficult” extremes. This aligns with HCI and cognitive psychology work on processing fluency, in which rapid responses signal low cognitive load and decisive task engagement, whereas prolonged responses reflect uncertainty, deliberation, or difficulty \cite{horwitz2017using, leipold2024detecting, Alter2009}. However, response time alone proved insufficient: when used in isolation, classification accuracy dropped sharply. This underscores that while response time captures task-oriented effort, it lacks sensitivity to the internal arousal dynamics that distinguish, for example, a stressed user performing an easy task from a relaxed user performing a difficult one. The importance of tonic EDA in distinguishing the mixed states (“relaxed \& difficult” vs. “stressed \& easy”) highlights precisely this nuance. Tonic EDA reflects sustained sympathetic activation, capturing stress levels even when behavioral performance remains fluent. This dissociation is particularly relevant for HCI: a respondent may complete an item quickly yet still report high stress, an internal state that behavioral measures alone would miss. The complementary roles of response time and tonic EDA thus map onto two fundamental components of cognitive–affective experience: effortful processing and physiological arousal, confirming that user experience must be evaluated along two axes: how hard the user's central executive system is working and the user's sustained internal stress level \cite{Antosz2016}. Effective adaptive systems must therefore consider both dimensions to avoid misinterpreting fluent behavior as comfort or slow behavior as distress. Beyond these primary indicators, SHAP values revealed that mouse trajectory distance contributed meaningfully to identifying the “stressed \& difficult” state. Larger or more erratic cursor movements during high-demand moments echo prior findings linking spatial inefficiency in input behavior to elevated cognitive load. Within our brief-interaction context, increased mouse distance likely reflects micro-level motor adjustments associated with simultaneous difficulty and stress, capturing a behavioral manifestation of cognitive–affective overload. The modeling and SHAP findings reinforce a broader implication for HCI: no single modality provides a complete picture of user state during rapid survey interactions. Behavioral features reflect task fluency; physiological features capture internal arousal; and trajectory dynamics reveal motor-level manifestations of combined stress and difficulty.

\subsection{Design and Practical Implications: A Tiered Framework for Adaptive Survey Systems}
Our findings reveal that behavioral and physiological signals provide complementary windows into users’ cognitive and affective states in brief interactions, but that design intent, subjective experience, and signal expression do not always align. Building on these mappings, and informed by our multimodal machine learning results, we propose a tiered micro-intervention framework for adaptive survey systems. The framework provides guidance on when and how real-time signals should trigger adjustments such as clarification, timing changes, or brief recovery opportunities.

\subsubsection{Match Intervention Type to Signal Modality} Behavioral and physiological channels signal different forms of struggle and therefore require different forms of support. Behavioral markers, mouse hesitations, hovers, complex trajectories, reflect cognitive difficulty and benefit from cognitive scaffolds such as clarifications or examples. Physiological markers, such as elevated tonic EDA, signal affective stress, suggesting the need for supportive interventions such as time-pressure removal or optional breaks. When both channels are elevated simultaneously, the system should offer meta-cognitive scaffolding such as confidence checks or the ability to flag items.
The differential utility of these signals also informs sensor selection across deployment contexts. For remote web surveys and large-scale online platforms, mouse tracking and response time provide substantial discriminatory power at effectively zero hardware cost, making them an ideal default for detecting cognitive struggle \cite{lenzner2010cognitive,leipold2024detecting}. Our results directly support this approach: behavioral signals showed high information density and strong sensitivity to difficulty. In high-stakes contexts where greater accuracy justifies additional sensing burden, physiological modalities such as EDA can reveal stress-related arousal patterns that behavioral signals alone may miss, enabled by smartwatch \cite{Lazarou2024, Siirtola2019} or webcam \cite{Wisiecka2022, Kraft2022}. Our findings showed tonic EDA was uniquely diagnostic of stress, while HR-related measures were comparatively less informative for brief interactions. Lightweight platforms may benefit from a hybrid model in which mouse-based inference serves all users, while optional wearables enhance detection for those with compatible devices.
These mappings follow from our results potentially showing that difficulty primarily affects behavioral hesitation, while time-pressure disproportionately affects physiological arousal.

\subsubsection{Validate Design Intent Against Subjective Experience} The divergences we observed between intended difficulty/stress and users’ self-reported experience underscore the need for systems that consider both subjective reports and task features. Because models trained on self-reported labels substantially outperformed those trained on experimental conditions, adaptive systems should periodically check subjective difficulty (“How was that question?”), compare presupposed difficulty with experienced difficulty at scale, and retrain detection models to reflect users’ lived experience. Pre-testing questions can also help predict subjective difficulty (e.g., with cognitive interviewing \cite{drennan2003CognitiveInterviewingVerbal}). This reduces the risk of adaptive behaviors that respond to designer assumptions rather than user reality.

\subsubsection{Support User Agency Through Transparent Adaptive Control} Because people vary widely in how they appraise stress and difficulty, adaptive systems must preserve user agency by making sensing and interventions transparent, controllable, and consent-driven. Rather than inferring user needs and acting unilaterally, systems should communicate which signals are being monitored and why (e.g., “We use mouse behavior to detect moments of hesitation”), and allow users to opt into specific adaptive behaviors such as stress-based break suggestions or automatic hint availability. When interventions are needed, they should be offered as choices rather than automatic system actions, e.g., “You’ve been on this question for 45 seconds. Would you like a hint, skip, or continue?”, allowing users to decide when and how support is used. After a session, systems can reinforce trust by providing a brief summary of how adaptive behaviors were triggered (“We suggested three breaks based on stress signals—was that helpful?”). This emphasis on transparency and granular control is essential when collecting sensitive behavioral or physiological data, where a one-size-fits-all adaptation strategy might misalign with individual differences in stress appraisal and risk undermining user trust.

\subsubsection{Intervene Through Tiered Escalation at Natural Task Boundaries} Adaptive survey systems should rely on tiered escalation rather than immediate reaction, responding only when signals of struggle persist across multiple questions and intervening at natural breaks in task flow. Because our per-item analyses show that cognitive and affective markers can fluctuate within short 5–17 second windows, intervening after a single spike risks misinterpreting momentary hesitation, distraction, or exploratory behavior as genuine difficulty. Mid-question interruptions can break concentration and distort the very behaviors the system aims to observe \cite{iqbalInvestigatingEffectivenessMental2005}.
Instead, systems should combine temporal escalation with boundary-aware timing. Systems should treat the first occurrence of elevated behavioral or physiological signals as a monitoring event only. Between questions—where support can be offered unobtrusively without interrupting cognitive flow—systems can deploy subtle scaffolds when hesitation or arousal patterns reappear over two or three consecutive items. These might include expanded definitions, optional tooltips, or slightly increased text spacing, framed as gentle prompts ("That looked challenging—would you like a hint on the next one?"). When signals continue to accumulate across four or more items, more explicit support becomes appropriate at the next question boundary, including offering examples, enabling "save for later," or suggesting short breaks.
Between blocks, the system can introduce higher-level adjustments such as modifying difficulty sequencing, relaxing timers, shortening the next block, or reducing question complexity—addressing patterns that reflect cumulative overload rather than item-specific struggle. End-of-session moments are appropriate for meta-cognitive interventions such as offering to review flagged questions or summarizing difficulty patterns. Systems should reserve in-flow interventions for only the most extreme cases, such as prolonged inactivity combined with strong hesitation markers.
This progression from passive monitoring to subtle scaffolding to explicit assistance to structural adaptation, delivered at natural task boundaries, prevents premature or intrusive interventions while still addressing sustained difficulty. By grounding adaptation in signal continuity rather than isolated deviations, and by timing support to align with the natural rhythm of users' task progress, designers can better distinguish transient variability from true cognitive-affective strain, supporting users without overstepping their autonomy or disrupting task flow. 

\subsection{Limitations and Future Work}
While our study offers valuable insights into the relationship between cognitive load, affective status, and multimodal behavioral and physiological measures, several limitations should be acknowledged, which also point to directions for future research. First, our data collection was conducted in a controlled lab environment, which, although useful for isolating variables, may limit the immediate transferability of our findings to a distributed data collection scenario. Real-world survey and assessment environments are more varied, multitasked, and interruption-prone. However, the modalities we analyze are increasingly accessible outside the lab: behavioral signals such as mouse movement and response timing are already available in virtually all online survey platforms \cite{horwitz2017using, horwitz2020learning}; physiological indicators such as EDA and HRV can be captured through common consumer wearables \cite{Lazarou2024, Siirtola2019}; and eye-movement proxies (e.g., blink rate, gaze dispersion) can be estimated through standard webcams \cite{Wisiecka2022, Kraft2022}. Major survey panels and commercial platforms have begun piloting integrations with smartwatch data, suggesting a realistic pathway for deploying adaptive mechanisms at scale \cite{Kapteyn2024}. We therefore see our findings not as limited to laboratory use but as a foundation for future real-world validation studies that exploit emerging, low-burden sensing technologies. Evaluating how these adaptive models perform under naturalistic noise conditions remains an important next step.

Second, our methodology involved collecting highly sensitive and private data (e.g., physiological signals, detailed behavior tracking), which raises privacy and ethical concerns that may complicate deployment in real-world applications. Addressing data security and participant consent will be essential for broader adoption. For example, information processing could be handled on the client side, making it unnecessary to transmit raw data.

Third, the demographic composition of our sample was biased toward younger and middle-aged adults, potentially limiting the generalizability of our findings to older populations, who may exhibit different cognitive and physiological patterns. Although 9 non-native speakers with intermediate to advanced proficiency participated in our experiment, this was not a central concern because of the within-subject design, and because the tasks were piloted for clarity and required only basic linguistic comprehension.

Moreover, in terms of data processing and modeling, we primarily relied on a set of basic, transparent and aggregated features (e.g., total counts, average durations) from physiological and behavioral data. Future work could benefit from incorporating richer feature sets, such as descriptive statistics (e.g., mean, standard deviation, min/max) or even raw signal data and complete mouse trajectories, which may capture diverse user state patterns. Then, while we evaluated the classification performance using three traditional algorithms, SVM, RF, and KNN, we did not explore more advanced or data-hungry models. With larger datasets and more granular input, future studies could implement deep learning architectures to better capture the temporal and non-linear dynamics in physiological and behavioral responses.

\section{Conclusion}
In this study, we investigated participants’ cognitive load and mental stress during multiple-choice question tasks, spanning both general knowledge and cognitive test items. A key contribution of our work lies in the structured data collection protocol, which combined wearable and ubiquitous sensors with fine-grained behavioral tracking to capture real-time responses at the level of individual question trials. By distinguishing between objectively defined task conditions and subjectively perceived states via immediate self-reports, we were able to examine not only the intended effects of task manipulations but also how individuals internally appraise stress and difficulty. This distinction is essential, as our results show that subjective experience can diverge meaningfully from experimental design, underscoring the methodological importance of jointly considering objective and subjective indicators.

Our predictive models demonstrate the feasibility of capturing meaningful cognitive-affective signals in highly dynamic, short-duration task contexts. More importantly, they reveal which behavioral and physiological features are selectively sensitive to different demands within sub–20-second windows, extending adaptive sensing beyond continuous tasks to discrete survey items. These insights highlight the complementary value of combining behavioral trajectories with physiological data to capture dimensions of user experience that single modalities or coarse task labels would overlook.

Overall, our work contributes not only an initial set of modeling results, which demonstrate the promise of multimodal behavioral and physiological signals for fine-grained inference in short web interactions, but also a validated protocol and empirical foundation for understanding how design factors and subjective appraisals jointly shape measurable signals in surveys. Building on these mappings, we further propose a tiered micro-intervention framework that outlines when and how real-time adjustments, such as clarifications, timing changes, or brief breaks, might be triggered in adaptive survey systems. Future research can build on this foundation by expanding the dataset, incorporating raw signal representations, and testing advanced modeling architectures. Ultimately, the most promising direction may be to use such data-driven insights to inform the design of adaptive interfaces, where the ability to detect and respond to user strain, even imperfectly, can already provide meaningful benefits.

\section{Open Science \& Transparency}
We invite readers to reproduce and build upon our findings. 
The experiment was preregistered, and the dataset and analysis scripts are open-sourced and available at this \href{https://osf.io/ehkjy/?view_only=f19c1065f8a348fb8073192818326099}{link} at the Open Science Framework (OSF). During the preparation of this work, the authors used OpenAI’s GPT-4o and Grammarly for grammar and style editing. All content was reviewed and edited by the authors, who take full responsibility for the final publication.
%
%



\bibliographystyle{ACM-Reference-Format}
\bibliography{bibliography}
\newpage

\appendix
\section{Descriptive Statistics of the Self-reported Ratings}
\label{sec:sdp}
\begin{figure}[htbp]
    \centering
    \includegraphics[width=0.67\textwidth]{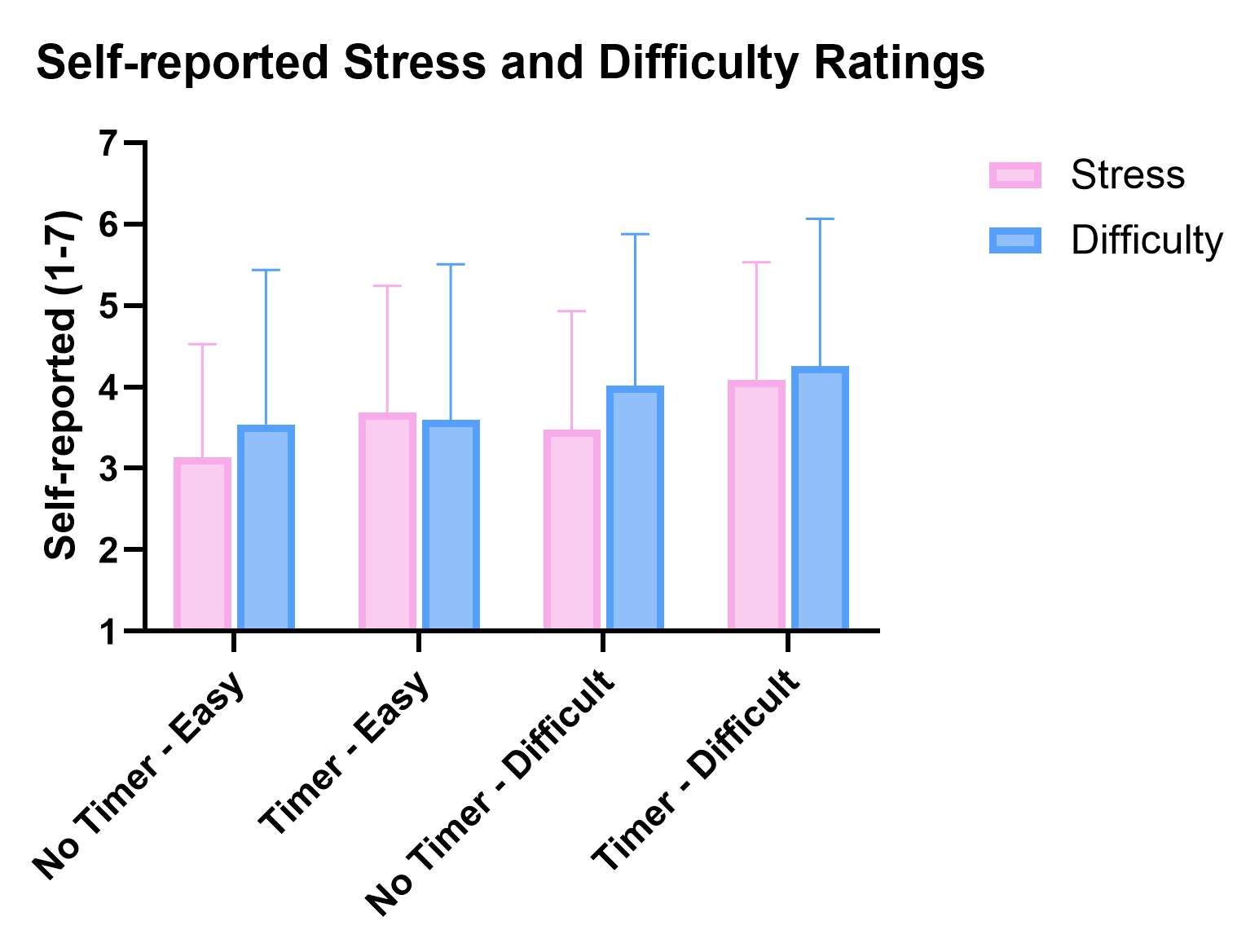}
    \Description{A plot showing the mean and standard deviation of self-reported ratings over manipulated conditions.}
    \caption{Mean and Standard Deviation of the self-reported ratings over manipulated conditions}
    \label{fig:sr}
\end{figure}

\section{Visualization of Self-reported Ratings}
\label{sec:vis}
\begin{figure}[htbp]
    \centering
    \begin{subfigure}[b]{0.45\textwidth}
        \includegraphics[width=\textwidth]{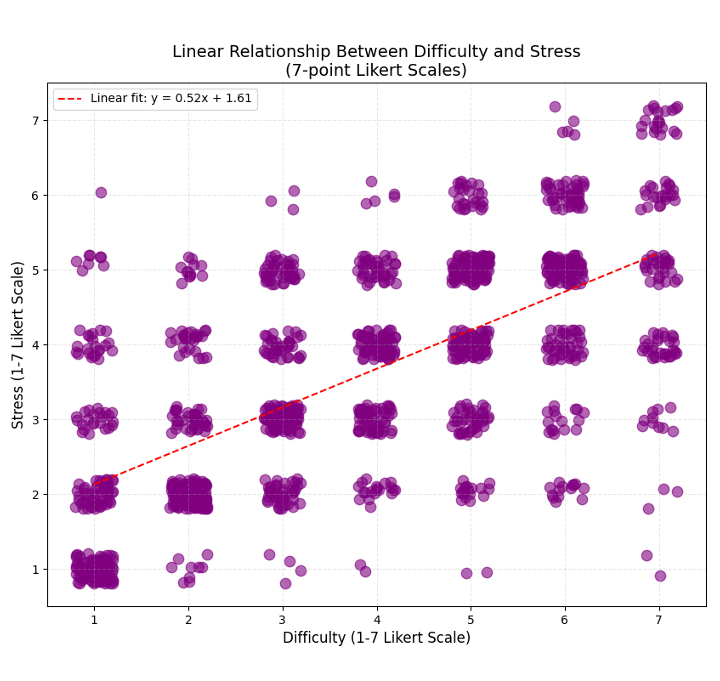}
        \caption{Linear relationship}
        \label{fig:linear}
    \end{subfigure}
    \hfill
    \begin{subfigure}[b]{0.45\textwidth}
        \includegraphics[width=\textwidth]{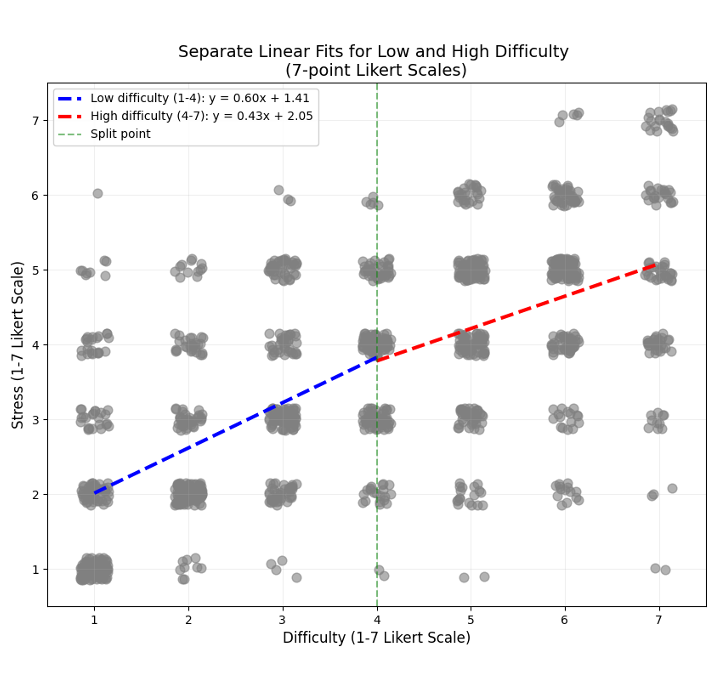}
        \caption{Split regression at difficulty = 4}
        \label{fig:split}
    \end{subfigure}
    
    \bigskip
    \begin{minipage}{\textwidth}
    \centering
    \caption{Visualization of self-reported stress-difficulty ratings. 
    (a) Simple linear regression shows the overall trend; 
    (b) Piecewise regression with split at difficulty = 4 (Neutral) demonstrates changing slopes across ranges. }
    \label{fig:combined}
    \end{minipage}
\end{figure}

\end{document}